\documentclass{article}

 \usepackage[final]{neurips_2025_ml4ps}

\usepackage[utf8]{inputenc} %
\usepackage[T1]{fontenc}    %
\usepackage{hyperref}       %
\usepackage{url}            %
\usepackage{booktabs}       %
\usepackage{amsfonts}       %
\usepackage{nicefrac}       %
\usepackage{microtype}      %
\usepackage{xcolor}         %
\usepackage{bm}

\usepackage{graphicx}
\usepackage{subcaption}

\usepackage{amsmath}

\title{Improving Posterior Inference of Galaxy Properties with Image-Based Conditional Flow Matching}

\author{%
  Mikaeel Yunus \\
  Department of Physics and Astronomy\\
  Johns Hopkins University\\
  Baltimore, MD 21218 \\
  \texttt{myunus1@jh.edu} \\
  \And
  John F. Wu \\
  Space Telescope Science Institute\\
  Baltimore, MD 21218 \\
  \texttt{jowu@stsci.edu} \\
  \And
  Benne W. Holwerda \\
  Department of Physics and Astronomy\\
  University of Louisville\\
  Louisville, KY 40208\\
  \texttt{benne.holwerda@louisville.edu} \\
}

\begin{document}

\maketitle

\begin{abstract}
  Estimating physical properties of galaxies from wide-field surveys remains a central challenge in astrophysics. While spectroscopy provides precise measurements, it is observationally expensive, and photometry discards morphological information that correlates with mass, star formation history, metallicity, and dust. We present a conditional flow matching (CFM) framework that leverages pixel-level imaging alongside photometry to improve posterior inference of galaxy properties. Using $\sim10^5$ SDSS galaxies, we compare models trained on photometry alone versus photometry plus images. The image+photometry model outperforms the photometry-only model in posterior inference and more reliably recovers known scaling relations. Morphological information also helps mitigate the dust--age degeneracy. Our results highlight the potential of integrating morphology into photometric SED fitting pipelines, opening a pathway towards more accurate and physically informed constraints on galaxy properties.
\end{abstract}

\section{Introduction}

Modern galaxy surveys capture millions of objects with unprecedented depth and detail, but turning these observations into reliable physical properties can be difficult. The gold standard is spectroscopy, where a galaxy's light is dispersed into a detailed spectrum across many wavelengths, and absorption and emission features pin down the physics of the galaxy with high precision. However, acquiring spectra for every galaxy in a wide survey is observationally expensive.\footnote{The advent of wide-field, complex spectroscopic surveys such as DESI \cite{DESI_Collaboration_2022, Dey_2019} and PFS \cite{greene2022} is rapidly increasing the availability of spectra. Nevertheless, spectroscopy will remain limited in scale relative to imaging, making photometric surveys the primary avenue for estimating galaxy properties at scale.}

Broadband photometry – measuring the integrated flux of a galaxy through a small number of wide filters – offers a cheaper and more scalable alternative for measuring galaxy properties. With only a relatively small number of flux measurements across different filters, stellar population synthesis (SPS) methods can recover many physical parameters with high accuracy \citep{Conroy_2013, iyer+2025}. However, this photometry-based approach discards valuable morphological information (e.g., spatial structure, color gradients, and overall visual appearance) that encodes properties such as stellar mass, star-formation history, and metallicity (e.g., \citep[][]{Wu_2019,Alfonzo+2024,Parker+2024}).

Since galaxy images contain a wealth of morphological information, it is natural to ask whether images can help improve galaxy property estimation. Recent work on this subject \cite{doorenbos_2024} has demonstrated that images can be used to generate optical spectra via generative models, from which galaxy properties can then be inferred. However, this approach relies on artificial spectra as an intermediate step before any galaxy physics can be constrained.

Alternatively, simulation-based inference (SBI) provides a framework to \textit{directly} incorporate galaxy morphology into physical property estimation (e.g., \cite{Cranmer+2020}). While SBI has been applied to galaxy property inference \cite{Alsing_2020,Hahn_2022}, only recently has imaging data been incorporated into such pipelines (e.g., \cite{iglesias-navarro}).

In this work, we test the hypothesis that explicitly including morphology from images in an SBI framework can sharpen galaxy property inference. To do so, we train two complementary models: a photometry-only baseline and a model augmented with latent representations of images. By comparing their posteriors, we directly assess the value of morphology for improving galaxy property estimates and potentially breaking astrophysical degeneracies.

\section{Data}

We use galaxies drawn from the Sloan Digital Sky Survey \citep{York_2000, Abazajian_2009} Main Galaxy Sample ($r<17.78$; \cite{Strauss_2002}). Starting from the SDSS \texttt{galSpecExtra} catalog of spectroscopically confirmed galaxies \citep{Kauffmann_2003, Brinchmann_2004, Tremonti_2004, Salim_2005}, we remove objects with unphysical values and discard systems with model magnitudes brighter than $r<16$ (see Figure~\ref{fig:galaxy-property-distributions}). All objects in our sample are bright, star-forming galaxies (based on BPT classification \cite{Baldwin}). After filtering, our working sample contains $106{,}800$ galaxies, which we split $80/10/10$ into training, validation, and test sets ($85{,}440/10{,}680/10{,}680$).

We work with five physical property variables: stellar mass $M_\star$, star formation rate (SFR), gas-phase metallicity $Z_{\rm{gas}}$, the narrow 4000\,\AA\ break index $D_n(4000)$ (a proxy for stellar population age), and the $V$-band dust attenuation $A_V$. Figure~\ref{fig:galaxy-property-distributions} shows the distributions of these properties in our sample.

We train models conditioned on SDSS \textit{ugriz} photometry and $128\times128$ image cutouts. We download these \textit{gri}-band cutouts from SDSS SkyServer at the native pixel scale ($0.396^{\prime\prime}\,\mathrm{pixel}^{-1}$). Each image is centered on the corresponding galaxy.

\section{Methods} \label{sec:methods}

We aim to infer posterior distributions of five galaxy properties – stellar mass, star formation rate, metallicity, $D_n(4000)$, and dust attenuation – from SDSS observations using Conditional Flow Matching (CFM) \citep{lipman2023flow, tong2024improving}. We train two complementary models: (i) a baseline that conditions only on \textit{ugriz} photometry, and (ii) a model that augments photometry with morphological information extracted from images. Throughout, we refer to these as the \textbf{photometry model} and the \textbf{image model}, respectively.

\textbf{Conditional Flow Matching.}
Let $\theta$ represent the galaxy properties and $\mathcal{D}$ the corresponding observational data (photometry and/or images). CFM learns a time-dependent velocity field $v_\phi(t,\theta,\mathcal D)$ that transports a simple prior to the posterior $p(\theta\mid\mathcal D)$. We use a linear interpolation path – i.e. we sample $\theta_0\!\sim\!\mathcal N(0,I)$, $t\!\sim\!\mathcal U(0,1)$, $\epsilon\!\sim\!\mathcal N(0,I)$ and form
\[
\theta_t=(1-t)\,\theta_0+t\,\theta_1+\sigma\,\epsilon,
\]
where $\theta_0$ is drawn from a Gaussian prior, $\theta_1$ denotes the target properties for each training example, and $\sigma$ is chosen to be $0.05$. This results in the probability path $p_t(\theta\mid \mathcal D,\theta_0,\theta_1)=\mathcal N\!\big(\theta;\,(1-t)\theta_0+t\theta_1,\;\sigma^2 I\big)$ \cite{gagneux2025}. We train with an MSE loss to the time-independent target velocity $\theta_1-\theta_0$. At test time we integrate $\dot\theta=v_\phi(t,\theta,\mathcal D)$ from $t{=}0$ to $1$ using the fourth-order Runge-Kutta method (RK4) with 100 steps, drawing 1000 trajectories per object to approximate the posterior.

\textbf{Architectures.}
For our \textbf{photometry} model, the velocity network is an MLP (three layers, width 256) whose input is the concatenation $[t;\theta;f_{\text{phot}}]$, where $f_{\text{phot}}\in\mathbb{R}^5$ is the normalized photometry. The input dimensionality is $1+d_\theta+5=11$ with $d_\theta=5$. For our \textbf{image} model, we encode a $128{\times}128$ RGB cutout with a CNN comprising four stride-1 convolutional blocks with average pooling, followed by global average pooling, yielding a 256-D representation $f_{\text{img}}$. We use average pooling to preserve extended light distributions, which are informative for galaxy property estimation. We then condition the velocity MLP (three layers, width 256) on $[t;\theta;f_{\text{img}};f_{\text{phot}}]$ (input size $1{+}d_\theta{+}256{+}5=267$). 

\textbf{Training Details.}
Models are implemented in PyTorch and trained with AdamW (learning rate $5\times10^{-5}$, batch size 64) with early stopping on validation loss. All continuous variables (photometry and properties) are standardized to zero mean and unit variance; inverse transformations are applied to samples for evaluation. We hold out a separate test set, used only for posterior evaluation. Training uses PyTorch \texttt{DataParallel} on four NVIDIA V100 GPUs.

\textbf{Posterior Performance Metrics. }
We evaluate our posteriors using two complementary statistics. 

\emph{Accuracy:} For each object we compute $\Delta \log p(\theta_\ast; \mathcal{D}) = \log p(\theta_\ast|\mathcal{D}) - \log p(\theta_\ast)$, where $\theta_\ast$ are the target parameters for that object, and $\mathcal{D}$ is the conditioning data. Positive values mean the posterior assigns higher density to the target than the prior does, i.e.\ a per-object Bayes factor gain. (Note that the ``prior'' here is the empirical distribution of properties in the spectroscopic dataset, not the Gaussian prior used in CFM training.)

\emph{Informativeness:} We also measure the Kullback-Leibler divergence $D_{\mathrm{KL}}\!\left[p(\theta|\mathcal{D})\,\|\,p(\theta)\right]$, quantifying how different the posterior distribution is from the prior distribution. Averaging $D_{\mathrm{KL}}$ over $\mathcal{D}$ yields the mutual information $I(\theta;\mathcal{D})$, a population-level summary of information gain.

\section{Results}

\begin{figure}[t]
    \centering
    \includegraphics[width=\linewidth, trim=0in 1in 0in 1in]{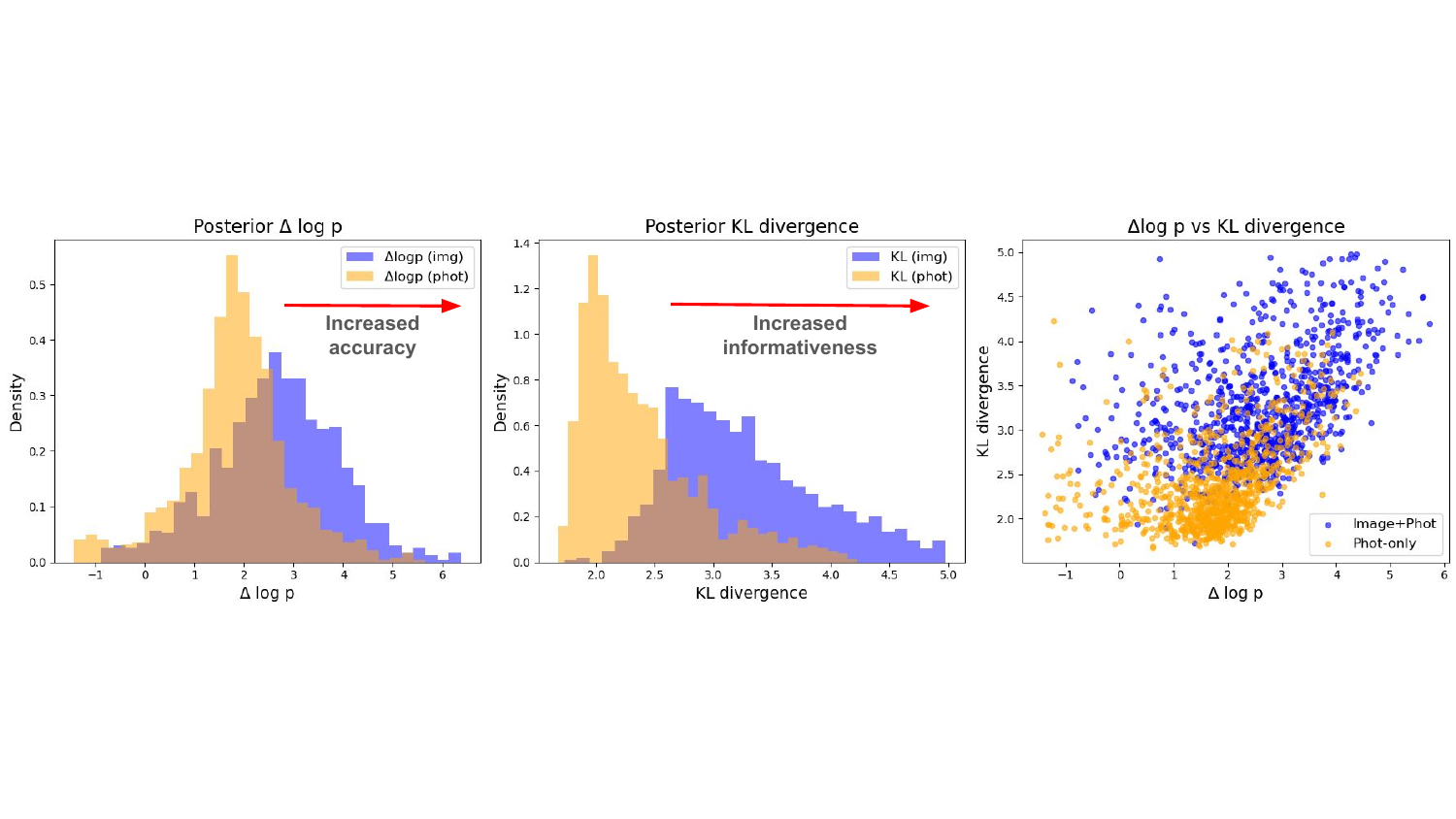}
        \caption{Left: distributions of $\Delta\log p(\theta_\ast;\mathcal D)$ for the image model (purple, $\mu=2.17, \ \sigma=3.30$) and the photometry model (yellow, $\mu=1.26, \ \sigma=3.98$). Middle: distributions of $D_{\mathrm{KL}}\!\left[p(\theta\!\mid\!\mathcal D)\,\|\,p(\theta)\right]$ for the image model (purple, $\mu=3.41, \ \sigma=0.95$) and the photometry model (yellow, $\mu=2.55, \ \sigma=0.97$). Right: per-object scatter plot of $\Delta\log p$ vs.\ $D_{\mathrm{KL}}$ (color indicates model). Panels show $N=1000$ galaxies from the test set. For these objects, the image model attains \textbf{higher $\Delta\log p$ for 81.5\%} of objects and \textbf{higher $D_{\mathrm{KL}}$ for 96.5\%} of objects compared to the photometry model.\vspace{-1em}
        }
    \label{fig:posterior-quality}
\end{figure}

We compare posterior quality between the \textbf{image} and \textbf{photometry} models using the two metrics defined in Section~\ref{sec:methods}: (i) the \emph{accuracy} statistic
$\Delta\log p(\theta_\ast; \mathcal D)$,
evaluated at the target $\theta_\ast$, and (ii) the \emph{informativeness} statistic
$D_{\mathrm{KL}}\!\left[p(\theta\!\mid\!\mathcal D)\,\|\,p(\theta)\right]$,
which measures departure from the empirical (dataset) prior. Figure~\ref{fig:posterior-quality} summarizes these comparisons.

The image model shifts $\Delta\log p$ rightward relative to the photometry baseline, indicating higher posterior density at the target for more galaxies (higher per-object Bayes-factor gain). Simultaneously, $D_{\mathrm{KL}}$ increases, showing that the image-conditioned posteriors move further from the dataset prior. The joint trend in the right panel demonstrates that informativeness and accuracy both increase: objects with larger $D_{\mathrm{KL}}$ typically also exhibit larger $\Delta\log p$. A small number of outliers with negative $\Delta\log p$ remain, but since these appear in both models, they likely reflect CFM architecture limitations rather than the inclusion of morphological information in the image model.

In addition to our per-object posterior accuracy metrics, we evaluate how well each model reproduces the \emph{population-level} distribution of physical parameters. Specifically, for each galaxy property variable, we compute the Wasserstein distance (WD) between (i) its one-dimensional marginal distribution in the test set and (ii) the distribution of posterior means obtained from each model across the test set. This metric evaluates whether the model recovers the global population distribution. Smaller distances indicate that the model's predictions, aggregated across the test set, match the true population more closely. Table~\ref{tab:wasserstein} reports these Wasserstein distances for both the image model and the photometry model.

\begin{table}[h!]
\centering
\begin{tabular}{lccc}
\toprule
\textbf{Property} & \textbf{WD (Image Model)} & \textbf{WD (Photometry Model)} & \textbf{Difference} \\
\midrule
$M_\star$     & 0.0264 & 0.0547 & 0.0283 \\
SFR & 0.0639 & 0.1119 & 0.0480 \\
$Z_{\rm{gas}}$ & 0.0156 & 0.0302 & 0.0146 \\
$D_n(4000)$      & 0.0103 & 0.0131 & 0.0028 \\
$A_V$              & 0.1937 & 0.2565 & 0.0628 \\
\bottomrule
\end{tabular}
\vspace{0.5em}
\caption{Wasserstein distances between the marginal test-set distribution of each variable and the distribution of posterior means predicted by the image model and the photometry model. Lower values indicate better agreement with the true population distribution. The rightmost column shows the absolute improvement (reduction in WD) achieved by adding morphology.}
\label{tab:wasserstein}
\end{table}

Across all parameters, the image model achieves substantially smaller Wasserstein distances than the photometry model, indicating that morphological information improves not only per-object posterior accuracy and informativeness, but also fidelity to population-level marginal distributions in the test set. 

\begin{figure}[h!]
    \centering
    \includegraphics[width=0.9\linewidth]{scatterplots.pdf}
        \caption{
        Image-model predictions (blue) recover known scaling relations in SDSS data (gray) more faithfully than the photometry model (green). Each row shows a different relation: $M_\star$--$Z_{\rm gas}$ (top), $M_\star$--SFR (middle), and SFR--$Z_{\rm gas}$ (bottom). Red boxes mark selections defined on the image-model predictions, with three representative galaxy cutouts shown on the right.}
    \label{fig:scatterplots}
\end{figure}

In Figure~\ref{fig:scatterplots}, we directly compare scaling relations (see \cite{Noeske, Daddi, Elbaz, Tremonti_2004, Lara-Lopez:2013sua}) recovered by the two models. 
The top, middle, and bottom rows show photometry-model and image-model predictions on our test set in the $M_\star$--$Z_{\rm gas}$, $M_\star$--SFR, and SFR--$Z_{\rm gas}$ planes, respectively. These predictions are obtained for each galaxy by drawing 1000 samples from the appropriate CFM model, finding the mean, and displaying it as a point in the appropriate scatterplot. In each case, the image model (blue) better reproduces known galaxy scaling relations represented by SDSS data (gray) than the photometry model (green). 

In each row of Figure~\ref{fig:scatterplots}, we also highlight selection boxes (red) defined on the image model predictions and display three galaxy cutouts from each selection. These examples show that the image model yields visually coherent samples aligned with astrophysical expectations – for instance, low-$M_\star$, low-SFR galaxies appear blue and diffuse, as seen in the middle row.

\section{Discussion}

We have developed an image-conditioned flow matching model that outperforms a photometry-only model, and more accurately reproduces known galaxy scaling relations. This supports our aim to leverage morphological information to constrain galaxy properties. 

\begin{figure}[t]
    \centering
    \includegraphics[width=0.8\linewidth]{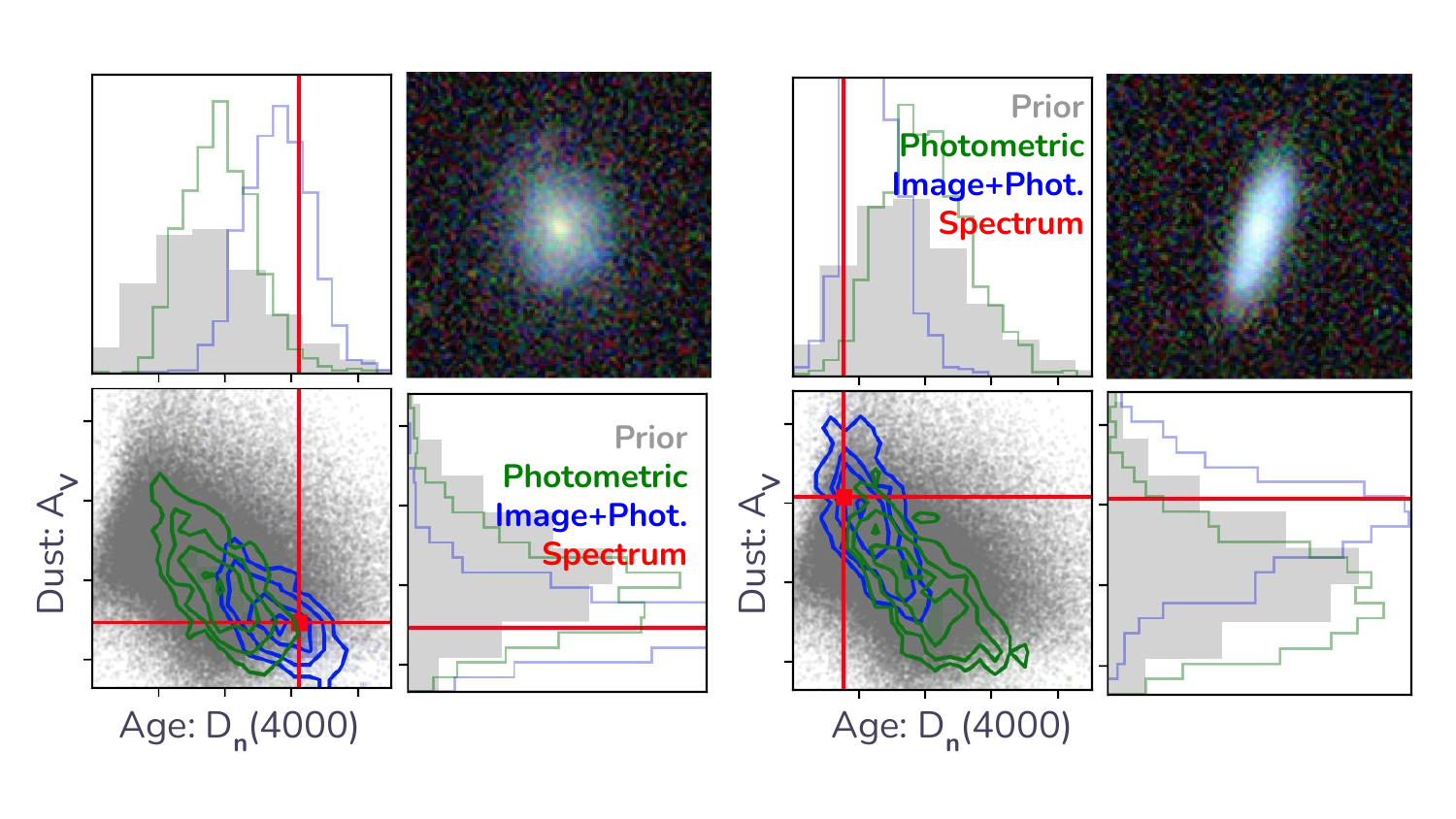}
        \caption{Corner plots of $A_V$ (dust attenuation) versus $D_n(4000)$ (stellar age proxy) for two representative galaxies. 
        \emph{Left:} old, dust-poor galaxy. \emph{Right:} young, dust-rich galaxy. 
        Contours show posterior predictions for the photometry model (green) and image model (blue); red crosses mark spectroscopic targets. 
        The image model moves closer to the spectroscopic target than the photometry model does in both cases, indicating potential to weaken the dust--age degeneracy.}
    \label{fig:image_vs_phot_corner}
\end{figure}

Long-standing degeneracies in galaxy physics, such as the dust--age degeneracy, provide another motivation for incorporating morphology into posterior inference of galaxy properties. Figure~\ref{fig:image_vs_phot_corner} depicts two illustrative cases: an old, dust-poor system (left) and a young, dust-rich system (right). In both examples, the image model (blue) shifts closer to the spectroscopic target (red) compared to the photometry-only model (green), suggesting that morphology can provide improvements even when $A_V$ is difficult to constrain. However, overall constraints on $A_V$ remain weak, and we therefore only partially disentangle $A_V$ from $D_n(4000)$. Nonetheless, Figure~\ref{fig:image_vs_phot_corner} highlights the potential of our approach to break dust--age degeneracies once integrated into a more comprehensive pipeline. 

Looking ahead, we plan to combine physics-based photometric SED fitting with our flow-based framework in order to incorporate morphological information into galaxy property inference. Traditional SED fitting encodes physics-based priors and captures parameter extremes well, but it has not yet offered a statistical pathway to leverage morphology. Our work can provide such a pathway. Future pipelines that merge SED models with SBI can retain the physical interpretability of SED fitting while using morphology to supply missing information, thereby yielding more accurate and comprehensive galaxy property estimates.


\bibliographystyle{plain}

\appendix
\renewcommand{\thefigure}{A.\arabic{figure}}
\setcounter{figure}{0}  %

\section{Appendix}

\begin{figure}[h] 
    \centering 
    \includegraphics[width = \textwidth]{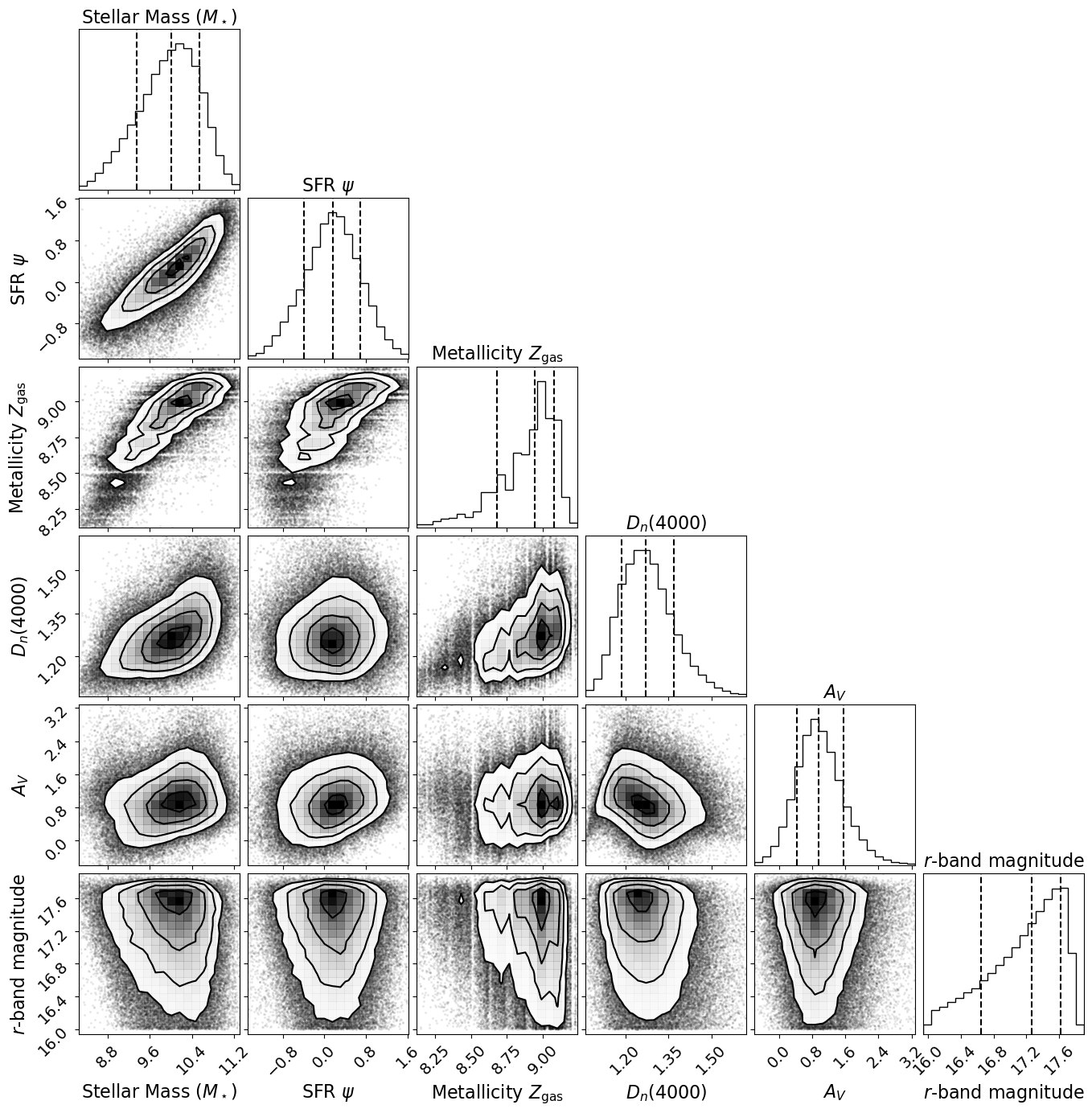}
    \vspace{-0.5em} 
        \caption{Distributions of SDSS galaxy properties and $r$-band magnitudes. The 16th, 50th, and 84th percentiles of the univariate distributions are labeled with dashed lines. 
    \vspace{-1em}} 
    \label{fig:galaxy-property-distributions} 
\end{figure}

Figure \ref{fig:galaxy-property-distributions} shows the univariate and bivariate distributions of SDSS galaxy properties and $r$-band magnitudes in our spectroscopic dataset.

\end{document}